\documentclass[aps,prb,twocolumn,floatfix]{revtex4}
\usepackage{color,graphicx,pstricks}
\usepackage{psfrag}
\usepackage{epstopdf}
\usepackage{textcomp}
\usepackage{hyperref}
\hypersetup{
  colorlinks=true,
  citecolor=blue,
  }
\usepackage{amsmath}
\begin{document}

\title {Magnetotransport Properties of Ferromagnetic/Antiferromagnetic Superlattices: Probing the role
of induced magnetization in antiferromagnetic layer}

\author{Sandip Halder$^1$, Snehal Mandal$^2$ and Kalpataru Pradhan$^1$\footnote{kalpataru.pradhan@saha.ac.in}}

\affiliation{$^1$Theory Division, Saha Institute of Nuclear Physics,
A CI of Homi Bhabha National Institute, Kolkata 700064, India\\
$^2$Centre for Quantum Engineering, Research and Education,
TCG Centres for Research and Education in Science and Technology,
Kolkata 700091, India.}

\begin{abstract}
We investigate the magnetic and transport properties of 
the $La_{1-x}Sr_{x}MnO_{3}$ (LSMO)/$Pr_{1-x}Ca_{x}MnO_{3}$ (PCMO) 
like ferromagnetic/antiferromagnetic superlattices in three dimensions 
using a two orbitals double exchange model incorporating the 
Jahn-Teller lattice distortions, superexchange interactions and 
long-range Coulomb interactions. In our simulations we primarily 
focus on periodic arrangement of $w_L$ planes of ferromagnetic LSMO 
and $w_P$ planes of antiferromagnetic PCMO manganites, and set 
$w_L$+$w_P$ = 10. The induced ferromagnetic correlations in the parent 
PCMO layer decreases monotonically with increasing the PCMO layer 
width $w_P$ at high temperatures for both half-doping ($n =0.5$) and 
off-half-doping ($n = 0.55$) scenarios. As we decrease the temperature 
further the induced ferromagnetic moments in PCMO layer disappears or 
decreases considerably at half-doping due to the onset of charge ordering 
in antiferromagnetic PCMO layers. Overall, the magnetization in PCMO 
layer decreases at low temperatures and the metal-insulator transition 
temperature of the LSMO/PCMO superlattices increases with increase of 
the PCMO layer width $w_P$, similar to the experiments. On the other 
hand, at off-half-doping, the induced ferromagnetic moment survives 
even at low temperatures due to weakened charge ordering in PCMO layer 
and interestingly, varies nonmonotonically with PCMO layer width, in 
agreement with experiments. The nonmonotonic trend of the conductivity 
of the superlattice with increase of the PCMO layer width $w_P$ shows 
an one-to-one correspondence between conductivity of the superlattice 
and the induced ferromagnetic moments in the PCMO layer. We highlight 
the key role of induced ferromagnetic moment in PCMO layer in 
analyzing the magnetotransport properties of the LSMO/PCMO superlattices.
\end{abstract}

\maketitle

\section{Introduction}
The surfaces and interfaces of artificially layered complex oxide
heterostructures frequently display new properties~\cite{Huang,Ngai,Hellman,Chen,Hwang}
that differ significantly from their bulk counterparts~\cite{Ohtomo,Takahashi,Zheng},
arising mainly due to abrupt changes in electronic
density~\cite{Garcia,Gibert,Ohtomo1,Hoffman}, lowering of the crystal symmetry~\cite{Hwang},
growth-induced strain and/or defects~\cite{Kourkoutis} at the interfaces.
Surprisingly, the physics that emerges at the interfaces are in some cases
absent in the bulk constituent materials~\cite{Ohtomo,Takahashi,Zheng}.
For example, an unexpected ferromagnetic metallic phase is realized at the
interface of Mott-insulator $LaTiO_3$ and the band insulator $SrTiO_3$~\cite{Ohtomo2,Seo}.
This conducting magnetic phase that emerges at the interface is due to the
electronic charge transfer from the Ti-$d$ band of $LaTiO_3$ to the Ti-$d$ band
of $SrTiO_3$~\cite{Okamoto}. Overall, the induced magnetism at the interface
controls the magnetotransport properties of the superlattice system~\cite{Bruno}.
Interfacial magnetism~\cite{Brinkman,Koida} and superconductivity~\cite{Reyren,Gozar}
(or even their co-existence~\cite{Dikin}), exchange bias~\cite{Yu,Gibert1,Zhou},
magnetoelectric coupling~\cite{Zubko,Liang,Bhoi,Gupta,Lawes} are few other interesting
examples of the unique phenomena that materializes at the interfaces in oxide
heterostructures. These unusual phenomena are of great technological
importance~\cite{Mannhart,Brivio,Yajima,Shen,Ramesh}, for instance in designing
next generation spintronic devices~\cite{Zutic,Tsymbal,Teresa,Ziese,Bibes}.

Among various complex oxides, doped rare-earth based perovskite manganites
(R$_{1-x}$A$_x$MnO$_3$ where R and A represents the rare-earth and alkaline-earth
elements, respectively)~\cite{Dagotto} are believed to be among the strong
candidates for various practical
applications~\cite{Panagio,Krivorotov,Julliere,Salafranca,Sun,Jo}, such as
spin-valves~\cite{Julliere,Salafranca}, magnetic tunnel junctions~\cite{Sun,Jo} and
magnetic sensors~\cite{Bibes1}. These Mn-based oxide compounds exhibit a
myriad of phases, e.g., ferromagnetism (FM), antiferromagnetism (AF), charge-ordered (CO),
orbital ordered (OO) phases, manifested mainly by the strong correlation among the
spin, charge, orbital and lattice degrees of freedom~\cite{Tokura,Tokura1,Dagotto1,Kajimoto}.
The ground state phase can be tuned by using different combination of R and A elements
and their concentration~\cite{Dagotto2,Chatterji}. Phase competition occurs at the
boundary of various phases and give rise to colossal magnetoresistance like effects
near optimal doping $x = 0.33$~\cite{Jin,Xiong}. In other words, engineering
electronic phase separation can be used as an effective tool to enhance the
magnetoresistance~\cite{Uehara,Fath,Pradhan3}.

Electronic phase separation, which represents the coexistence of two competing
phases- for example an antiferromagnetic insulating (AF-I) phase with a
ferromagnetic metallic (FM-M) phase is one of the interesting property in
manganites~\cite{Moreo,Khomskii,Simon}. Upon application of a magnetic field,
the AF-I phase diminishes whereas the FM-M phase grows in size~\cite{Helmolt,Tokura,Tokura1},
giving rise to the intriguing phenomena- known as magnetoresistance~\cite{Helmolt,Tokura,Tokura1,Jin,Xiong}.
Generally, half doped ($x$ = 0.5) manganite systems depicts charge ordered
antiferromagnetic phase which are highly insulating in nature~\cite{Tomoika,Kuwahara}.
Nanoscale electronic phase separation can be achieved in these kind of charge
ordered antiferromagnetic insulating systems by introducing ferromagnetic phase
fraction as follows: (i) by doping the system away from $x$ = 0.5~\cite{Kajimoto,Simon};
or (ii) by artificially making FM/AF heterostructures~\cite{Nieb1,Nieb2} and thus
inducing a ferromagnetic metallic clusters in an antiferromagnetic manganite. That
is to say, the phase competition can be engineered by preparing superlattices of
two different manganites where competition between the two phases across the
interfaces plays a vital role. Recently, a great amount of effort has been
devoted to engineer the electronic and magnetic phase competition in manganite
heterostructures.

The phase competition is mostly limited to few atomic layers near the interface.
For example, a ferromagnetic phase appears only at the interface in the manganite
superlattice $((LMO)_{2n}/(SMO)_{n})_{m}$~\cite{Bhattacharya,Adamo} due to
electronic phase competition although the individual LMO and SMO are antiferromagnetic
with Neel temperature 140 K and 260 K, respectively. The total system turns out
to be ferromagnetic for $n\leq 2$. A metal-insulator transition occurs for $n\geq 3$.
Resulting ferromagnetic metallic state at the interface is analogue to the solid
solution $La_{2/3}Sr_{1/3}MnO_{3}$. The magnetic properties are largely modified
at the interfaces due to the charge transfer across the interface~\cite{Zhong,Zhou,Gibert}.
On the other hand, a C-type AF phase is observed in $((LMO)_{n}/(SMO)_{2n})_{m}$
superlattice~\cite{May} for n = 1 and n = 2 which is equivalent to the properties
of the bulk $La_{1/3}Sr_{2/3}MnO_{3}$. Interestingly, the antiferromagnetic Neel
temperature increases by $\sim 70 K$ in superlattices as compared to the solid
solution. In general, phase competition emerging from the charge transfer from LMO
to SMO at the interface for n = 1 and n = 2 guides the LMO/SMO superlattice systems
to retain the magnetic phases which are obtained in the counterpart solid
solution.

The synthetically created nanoscale electronic phase competition is also reported
in manganite superlattices prepared using ferromagnetic manganites (like
$La_{2/3}Sr_{1/3}MnO_{3}$ and $La_{2/3}Ca_{1/3}MnO_{3}$) and charged ordered
antiferromagnetic insulating manganites (like $Pr_{2/3}Ca_{1/3}MnO_{3}$)~\cite{Mukhopadhyay,Li}
The charge transfer across the interface from LSMO to PCMO induces magnetization
in the antiferromagnetic PCMO layers. Amount of these induced ferromagnetic moments
depend on the width of the PCMO layer. Interestingly, it was observed that the induced
magnetization depicts a nonmonotonic behavior with variation of the PCMO layer
thickness~\cite{Nieb1,Nieb2} and the maximum magnetization is obtained at a thickness
comparable to the size of ferromagnetic nanoclusters as evidenced from polarized
neutron reflectivity~\cite{Saurel,Mercone}. In other words, when the dimension of
the induced ferromagnetic nanoclusters and the thickness of the PCMO layer are
comparable to each other one observes the maximum number of induced ferromagnetic
nanoclusters in PCMO layers. In an small external magnetic field, the magnetization
of induced ferromagnetic nanoclusters in PCMO layer orient along the field direction,
resulting in a decrease in spin scattering at the interfaces. So the mobility of the
carriers and also the conductivity of the system increases due to these induced
ferromagnetic nanoclusters. In fact, the magnetoresistance of the system follows
the same nonmonotonic behavior of the induced magnetization with the PCMO layer
width~\cite{Nieb2}. So, overall the width of the PCMO layer serves as a key
parameter to tune the magnetoresistance in LSMO/PCMO or LCMO/PCMO
superlattices ~\cite{Mukhopadhyay,Li,Cheng} and thus making the system
technologically important for applications in magnetic field sensors
and memory devices~\cite{Parkin}. Mostly, off-half-doped antiferromagnetic
manganites are used in the above studies.

The robustness of the charge and orbital ordered AF phase, which is mainly stable
at half-doping, gets diluted at off-half-doping~\cite{Urushibara,Tomoioka}. This is
why it seems that it is easier to induce ferromagnetic moments in off-half-doped
$Pr_{2/3}Ca_{1/3}MnO_{3}$ like antiferromagnetic systems when joined with LSMO
layers. Can one induce ferromagnetic moment in half-doped intermediate-bandwidth
$Pr_{1/2}Ca_{1/2}MnO_{3}$ manganites? Apparently, electronic phase separation has
also been reported in superlattices of FM $La_{1/2}Sr_{1/2}MnO_{3}$ and
AF $Pr_{1/2}Ca_{1/2}MnO_{3}$ manganites~\cite{Nakamura,Wadati}. The degree of
phase separation was tuned by variation of the individual layer width. The
electronic phase competition is more fascinating for
[$La_{1/2}Sr_{1/2}MnO_{3}$ (5 layers)/$Pr_{1/2}Ca_{1/2}MnO_{3}$ (5 layers)]$_{15}$
heterostructures at high temperatures (but, below the ferromagnetic
$T_C$ of $La_{1/2}Sr_{1/2}MnO_{3}$). In fact, a homogeneous ferromagnetic
metallic state arises at high temperatures, in contrast to the phase separated
state at low temperatures~\cite{Wadati}.

In this article, we have studied the magnetotransport properties of LSMO/PCMO
like FM/AF superlattices at half-doping ($n = 0.5$) and off-half-doping ($n = 0.55$)
to shed lights on the induced ferromagnetism in PCMO layers that give rise to
intriguing phase coexistence in these systems. We outline the one-to-one
correspondence between the magnetic and transport properties for various width
of PCMO layers using our Monte Carlo simulations. We organize the article in the
following way: In Sec.~\textbf{II}, we sketch the model Hamiltonian for manganite
based LSMO/PCMO (FM-M/AF-I) superlattice systems in three dimensions and briefly
discuss the methodology to solve the Hamiltonian. In addition, we present the
phase diagram of the bulk manganite system at half-doping ($n = 0.5$) in
the same section. In Sec.~\textbf{III}, we present suitable parameter set that
we use to construct individual ferromagnetic metallic and antiferromagnetic
insulating layers to represent LSMO/PCMO superlattices. Different measurable
physical observables are also highlighted in same section. We present the
magnetotransport properties of superlattices in Sec.~\textbf{IV} (for
half-doping case) and  Sec. \textbf{V} (for off-half-doping cases). In
Sec. \textbf{VI} we summarize our results.

\section{Model Hamiltonian and Method}
We use following two-orbital double exchange model~\cite{Dagotto2,Dong}
for itinerant $e_{g}$ electrons on a 3D cubic lattice:

\[H=\sum_{<ij>\sigma}^{\alpha \beta}(t_{\alpha \beta}^{ij}c_{i\alpha \sigma}^{\dagger}c_{j\beta \sigma}+H.c.)
-J_{H}\sum_{i}{\bf{S}}_{i}.{\bf{\sigma}}_{i}+J\sum_{<ij>}
{\bf{S}}_{i}.{\bf{S}}_{j}\]
\[-\lambda \sum_{i}{\bf{Q}}_{i}.{\bf{\tau}}_{i}+\frac{K}{2}\sum_{i}{Q_{i}}^{2}
-\mu \sum_{i}n_{i}\]

\noindent
Here, $c$ and $c^{\dagger}$ are the annihilation and creation operators
for itinerant $e_{g}$ electrons, respectively and $\alpha$, $\beta$ are
summed over the two $Mn-e_{g}$ orbitals $d_{3z^{2}-r^{2}}$ and $d_{x^{2}-y^{2}}$
(which are labeled as a and b). $t_{\alpha \beta}^{ij}$
implies the hopping amplitudes between $e_{g}$ orbitals of the nearest
neighbor sites which takes the form $(t_{aa}^{x}=t_{aa}^{y}\equiv t,
t_{bb}^{x}=t_{bb}^{y}\equiv t/3, t_{ab}^{x}=t_{ba}^{x}\equiv -t/\sqrt{3},
t_{ab}^{y}=t_{ba}^{y}\equiv t/\sqrt{3}, t_{aa}^{z}=t_{ab}^{z}=t_{ba}^{z}
\equiv 0, t_{bb}^{z}=4t/3)$~\cite{Dagotto,Dagotto2} (x, y, and z are represent the spatial
direction of the 3D cubic lattice). We set $t_{bb}^{z}$ = 1.1$\times$4$t$/3 for
our superlattice calculations which are often grown in $z$-direction.
The itinerant quantum $e_{g}$ spin (${\bf{\sigma}}_{i}$) is locally coupled
to the $t_{2g}$ spin (${\bf{S}}_{i}$) via Hund's coupling $J_{H}$. $J$ is the
AF superexchange interaction between the $t_{2g}$ spins on the nearest neighbor sites.
$\lambda$ denotes the strength of the John-Teller (JT) coupling between the distortion
${\bf{Q}}_{i}\equiv (Q_{ix},Q_{iz})$ and the orbital pseudospin ${\bf{\tau}}_{i}^{\mu}
=\sum_{\sigma}^{\alpha \beta}c_{i\alpha \sigma}^{\dagger}\Gamma_{\alpha
\beta}^{\mu}c_{i\beta \sigma}$. The stiffness constant of the JT modes
is denoted by $K$ (set equal to 1) and $\mu$ is the chemical potential of the system.
We set $t_{aa}=t=1$ and select as the reference energy scale. All the
parameters are measured in units of $t$. The JT distortions and the
$t_{2g}$ core spins are treated as classical~\cite{Dagotto3,Green}
variables and we set $|{\bf{S}}_{i}|=1$.

In manganite the estimated value of $t$ is around 0.2-0.4 eV~\cite{Popovic}
whereas the approximate value of the Hund's coupling strength is of the
order of 2 eV~\cite{Okimoto} i.e. one order larger than $t$. For this
reason we adopt the limit $J_{H}/t \rightarrow \infty $~\cite{Dagotto2}.
In this limit, the $e_{g}$ electron spin perfectly aligns along the
local $t_{2g}$ spins. Transforming the fermionic operators to this
local spin reference frame leads to the following spinless model for
the $e_{g}$ electrons:

\[H=\sum_{<ij>}^{\alpha \beta}({\tilde{t}}_{\alpha \beta}^{ij}c_{i\alpha}^{\dagger}
c_{j\beta}+H.c.)+J\sum_{<ij>}{\bf{S}}_{i}.{\bf{S}}_{j}-\lambda \sum_{i}{\bf{Q}}_{i}.{\bf{\tau}}_{i}\]
\[+\frac{K}{2}\sum_{i}{Q_{i}}^{2}-\mu \sum_{i}n_{i}\]

\noindent
where 
${\tilde{t}}_{\alpha \beta}^{ij}$ is defined as ${\tilde{t}}_{\alpha \beta}^{ij}=
\Theta_{ij} t_{\alpha \beta}^{ij}$ with $\Theta_{ij}=cos(\frac{\theta_{i}}{2})
cos(\frac{\theta_{j}}{2})+sin(\frac{\theta_{i}}{2})sin(\frac{\theta_{j}}{2})
e^{-i(\phi_{i}-\phi_{j})}$.
This factor controls the magnitude of the hopping amplitude depending on the
orientation of the $t_{2g}$ spins at site $i$ and $j$. Here, $\theta_{i}$
and $\phi_{i}$ represent the polar and azimuthal angles for the spin ${\bf{S}}_{i}$.
$c_{i\alpha}$ ($c_{i\alpha}^{\dagger}$) represents the fermion annihilation
(creation) operator at site $i$, in the orbital $\alpha$ with spin parallel
to ${\bf{S}}_{i}$. A Zeeman coupling term $H_{mag}=-h\sum_{i}S_{i}^{z}$ is
added to the Hamiltonian where the spin interacts with the external applied
magnetic field.

A long range Coulomb part $H_{lrc}=\sum_{i}\varphi_{i}n_{i}$ is added to the
Hamiltonian to control the amount of charge transfer across the interface of
FM-M/AF-I superlattices. $\varphi_{i}$ is the long range coulomb potential
which is determined self-consistently by setting the potential at the mean field
level~\cite{Okamoto1,Yunoki,Salafranca,Pradhan1}
as \[\varphi_{i}=\alpha t\sum_{j\neq i}\frac{<n_{j}>-Z_{j}}{|R_{i}-R_{j}|} \]
\noindent
where $Z_{j}$ is the charges of the fixed background ions (confined at the
Mn sites). $<n_{j}>$ is the density of the itinerant electrons in the Mn site
at $R_{j}$. The strength of the Coulomb interaction is determined by the
parameter $\alpha$ which is defined as $\alpha = e^{2}/\epsilon a t$,
$\epsilon$ and $a$ are the dielectric constant and lattice parameter
respectively. For the calculation of the manganite system in 3D the value
of $\alpha$ is generally chosen in the range of 0.1-1~\cite{Lin,Nanda,Yu1}.
We set $\alpha$=0.5 for our calculations.

\begin{figure}[!t]
\centering
\includegraphics[scale=0.35]{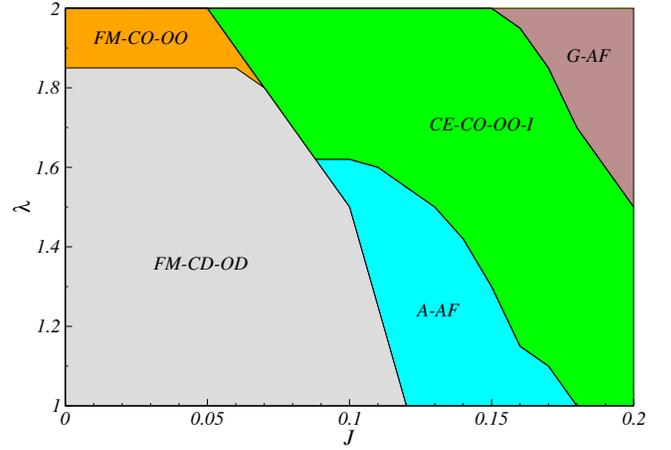}
\caption{
\label{fig01}
The ground state phase diagram ($T = 0.01$) of the bulk manganite systems at
electron density $n = 0.5$ for varying $J$ and $\lambda$ in three dimensions.
For details of the phases, please see the text.
}
\end{figure}

We anneal the classical variables by starting with the random configurations
at high temperatures. We diagonalize the itinerant electrons in different configurations
of classical $t_{2g}$ spins $\bf S$ and the lattice distortions $\bf Q$ in our
Monte Carlo calculations. During the Monte Carlo update in each temperature we use
2000 MC system sweeps. In each system sweep we visit each lattice site sequentially and
update the classical variables by Metropolis algorithm. To access large system size,
we use travelling cluster approximation (TCA) based Monte Carlo
technique~\cite{Kumar,Halder,Pradhan,Pradhan1}. In our calculations we access system
size of dimension $8\times8\times20$ by using TCA cluster $4\times4\times8$ to study
FM/AF superlattices. Long range coulomb potential $\varphi_{i}$ in $H_{lrc}$ is
solved self-consistently at every 10 steps interval during system sweeps
by checking the convergence of the electron density at each site.

In order to represent the constituent LSMO and PCMO layers correctly in LSMO/PCMO 
(FM/AF) superlattices we first briefly outline the ground state phase diagram
of the bulk system at electron density $n = 0.5$ (half-doping) by varying the
antiferromagnetic superexchange $J$ and Jahn-Teller coupling strength $\lambda$
in Fig.~\ref{fig01}. This bulk phase diagram is prepared using $8\times8\times8$
system size. To best of our knowledge, there has not been any study of the ground
state phase diagram at $n = 0.5$ in 3D systems. For smaller values of $J$ and
$\lambda$, the system remains in a double exchange dominated ferromagnetic state
to optimize the kinetic energy gain. Electron density of this ferromagnetic system
remains homogeneous due to the orbital disordering (OD) and charge disordering (CD).
As a result the system is metallic with finite density of state at the Fermi level
($\varepsilon_{F}$)~\cite{Pradhan}. As we increase the $\lambda$ value, keeping
the superexchange $J$ small, the magnetic state remains in ferromagnetic, but it
transits to a charge and orbital ordered insulating state (CO-OO-I) for $\lambda \geq 1.85$.
On the other hand, as superexchange $J$ is increased (keeping the $\lambda$ small)
the system transits to an A-type antiferromagnetic (A-AF) state (with structure
factor $S(\bf{q})$ peak at $\bf{q}$ = $(0, 0, \pi))$. We obtained a CE-type
antiferromagnetic, charge order (CO), orbital order (OO) phase (with simultaneous
$S(\bf{q})$ peaks at $\bf{q}$ = $(\pi, 0, \pi)$, $(0, \pi, \pi)$ and
$(\frac{\pi}{2}, \frac{\pi}{2}, \pi))$ by further increasing of the $J$ value.
We also obtain CE-CO-OO phase for intermediate value of $J$ and large value of $\lambda$.
This CE-CO-OO phase is strongly insulating due to the opening of gap at the
Fermi level~\cite{Pradhan,Aliaga,Popovic}. For large values of $J$ and $\lambda$
the phase changes to G-type AF (with structure factor $S(\bf{\pi, \pi, \pi})$)
which is shown in the top right hand corner of the phase diagram. Our phase
diagram is consistent with previous 2D results~\cite{Pradhan2,Yunoki1}.

\begin{figure}[!t]
\centering
\includegraphics[scale=0.36]{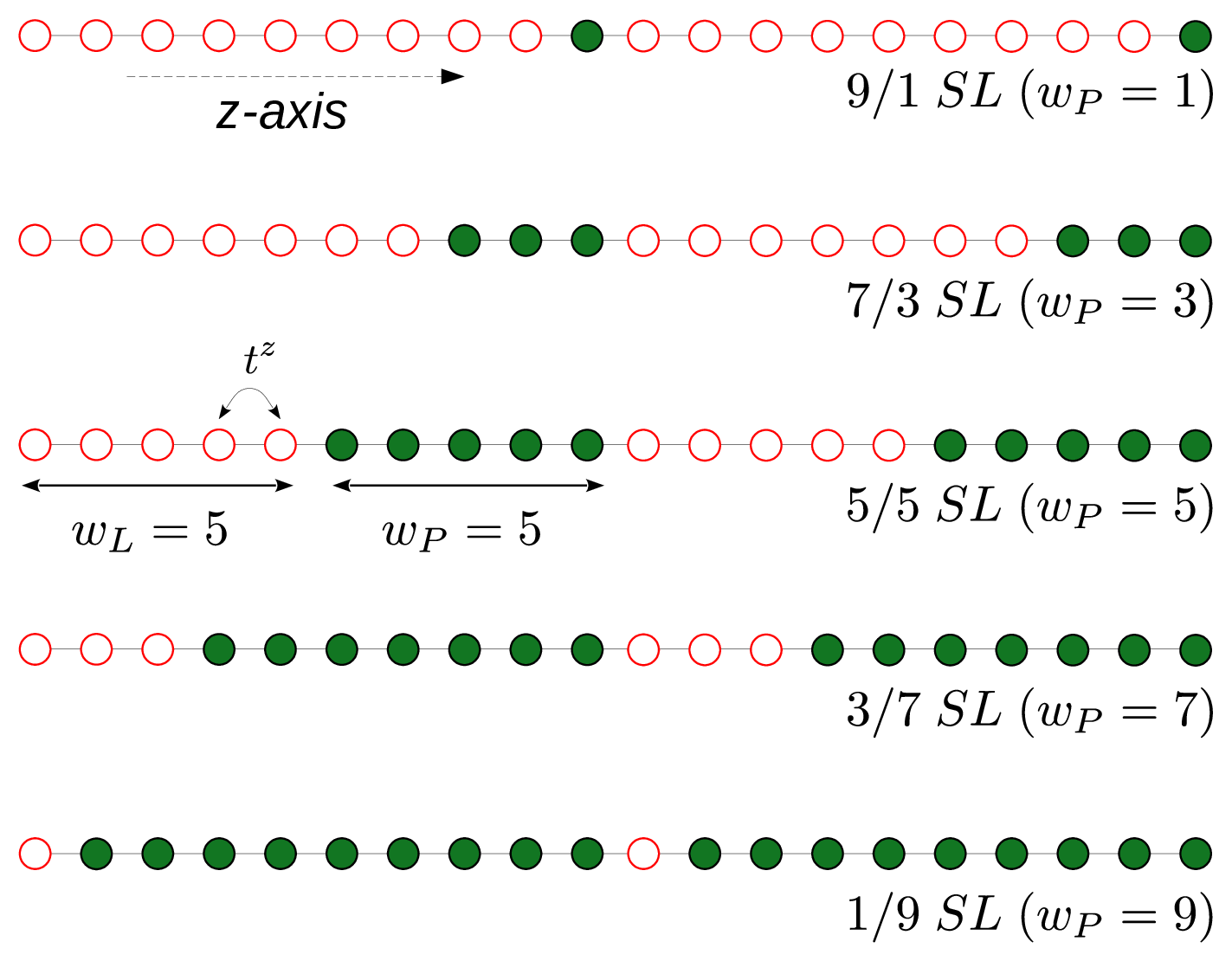}
\caption{
\label{fig02}
Schematic diagram of various LSMO/PCMO superlattices where each red open circles
(green solid circles) represents a $8\times8$ 2D $xy$ plane. $w_{L}$ ($w_{P}$) denotes
the width of the LSMO (PCMO) layer. Total thickness of the superlattices along the
z-direction is fixed to $L_{z} = 2\times (w_{L} + w_{P}) = 20$. $t^{z}$ is the
nearest neighbor hopping parameter along the z-direction. For details please see
the text. Periodic boundary condition is imposed in all three directions.
}
\end{figure}

{\section{Parameter sets to represent LSMO and PCMO materials}

To investigate the properties of LSMO/PCMO SLs we need to construct periodic
arrangement of LSMO (ferromagnetic) and PCMO (antiferromagnetic) layers. We assign
same electron density ($n$) to both the layers in our study. First we determine two
sets of parameters that we will assign to LSMO-like and PCMO-like materials to mimic
the essential physics of the individual constituent layers. From the phase diagram
(see Fig.~\ref{fig01}) it is clear that one can choose multiple parameter sets to
simulate a ferromagnetic metal (LSMO-like). Such a dilemma also exist for replicating
an antiferromagnetic insulating (PCMO-like) material. To proceed we set $J = 0.1$
for both LSMO and PCMO layers and vary $\lambda$ to differentiate the two systems.
For $\lambda = \lambda_{FM}\equiv 1.2$ ($\lambda = \lambda_{AF}\equiv 1.7$) the bulk
system is in FM-M (CE type AF-I) state as shown in Fig.~\ref{fig01} at $n = 0.5$.
In contrast, at off-half-doping case (say $n$ = 0.55-0.65) the long-range charge
and orbital ordered CE phase is diluted for $\lambda = 1.7$. But, the overall AF-I
nature of the system remains intact with local charge/magnetic ordering. The
ferromagnetic ordering for $\lambda=1.2$ at low temperatures is unaltered for
$n = 0.55-0.65$. So, we set $J$ = 0.1 and $\lambda = 1.2$ ($J$ = 0.1 and $\lambda = 1.7$)
to mimic LSMO-type (PCMO-type) layers in our calculations unless otherwise mentioned.
Henceforth, we call them LSMO and PCMO in our analysis.

We construct various LSMO/PCMO (FM-M/AF-I) superlattices [schematically shown
in Fig.~\ref{fig02}] where LSMO layer of width $w_L$ and PCMO layer of width
$w_P$ are periodically arranged. We employ periodic boundary conditions in our
calculations. The width of the LSMO and PCMO layers are adjusted in such a way
that the total thickness of the superlattice remains same along the out-of-plane
$z$ direction. We mentioned earlier that our system size is $8\times8\times20$ and
as a result total width of two LSMO layers and two PCMO layers is restricted to
20 planes. This construction is very similar to one of the reported experimental
setup where total thickness of LSMO and PCMO layers is fixed but thickness of
individual layers are varied~\cite{Nakamura}. These LSMO and PCMO layers are
coupled at the interface via hopping parameters ($t^{z}$) and superexchange
interactions ($J = 0.1$) across the interface. In this setup our primary focus
is to explore the induced magnetization in the antiferromagnetic PCMO layer with
variation of $w_P$ (width of the AF layer) and its effect on the out-of-plane
conductivity of the whole FM/AF superlattice systems. The strength of the long
range Coulomb interaction is chosen to be $\alpha = 0.5$ throughout our calculations.
All the calculations are performed in the presence of small applied magnetic field
$h = 0.005$ to align both FM layers in same direction.

We have calculated various physical observables to analyze the magnetotransport
properties of the superlattices. The average magnetization of the system is determined
by calculating $M = \frac{1}{N}\langle\sum_{i}S_{i}^{z}\rangle$ where N is the total
number of sites. The angular brackets represent the thermal average over the
Monte Carlo generated equilibrium configurations. In addition, all the physical
observables are averaged over ten different initial configurations of classical
variables. We also calculate the magnetization of the individual planes by using the
formula $M (i) = \frac{1}{L_{x}L_{y}}\langle\sum_{j\in i}S_{j}^{z}\rangle$, 
where $M (i)$ is the magnetization of the $i^{th}$ plane and $L_{x}L_{y}$ represents
number of sites in each plane. We also measure the local staggered magnetization
$M_{s} = \frac{1}{L_{x}L_{y}}\langle(\sum_{i\in A}S_{i}^{z}-\sum_{j\in B}S_{i}^{z})\rangle$
of individual planes. In addition, the onset of charge ordering in antiferromagnetic
PCMO layer is measured by calculating the staggered charge order
$CO_{s} = \frac{1}{N}\langle(\sum_{i\in A}n_{i}-\sum_{j\in B}n_{j})\rangle$.
To probe the induced magnetization (local ferromagnetic ordering)
in the antiferromagnetic PCMO layer we calculate the volume fraction of the FM
nanoclusters [V(SO)] which is defined as, fraction of sites, say $i$, for which
${\bf{S}}_{i}.{\bf{S}}_{j}\ge 0.5$ for all the six nearest neighbor sites $j$.
We also calculate the density of states (DOS) at frequency $\omega$ by using
the formula,  $DOS(\omega) = \frac{1}{N}\sum_{n}\delta(\omega-\epsilon_{n})$
where $\epsilon_{n}$ are the single particle eigen values.

To analyze the transport properties we calculate the conductivity of the
FM/AF superlattices in the $dc$ limit along the out-of-plane $z$ direction by
using Kubo-Greenwood formalism~\cite{Mahan,Kumar1,Chakraborty}:
\[\sigma(\omega)=\frac{A}{N}\sum_{\alpha, \beta}(n_{\alpha}-n_{\beta})
\frac{{|f_{\alpha \beta}|}^{2}}{\epsilon_{\beta}-\epsilon_{\alpha}}
\delta[\omega-(\epsilon_{\beta}-\epsilon_{\alpha})]\]

\noindent
where $A=\pi e^{2}/\hbar a$ (a is the lattice parameter). $f_{\alpha \beta}$
represents the matrix elements of the paramagnetic current operator
${\hat{j}}_{z}=i\sum_{i\alpha \beta}
(t_{\alpha \beta}^{z}\Theta_{i,i+z}c_{i,\alpha}^{\dagger}c_{i+z,\beta}-
{t_{\alpha \beta}^{z}}^{*}{\Theta_{i,i+z}}^{*}c_{i+z,\beta}^{\dagger}c_{i,\alpha})$
between the eigen states $|\psi_{\alpha}>$ and $|\psi_{\beta}>$ with corresponding
eigen energies $\epsilon_{\alpha}$ and $\epsilon_{\beta}$, respectively
and $n_{\alpha} = \theta(\mu - \epsilon_{\alpha})$. We extract
$dc$ conductivity by calculating the average  conductivity for a small low-frequency
interval $\Delta \omega$ defined as \[ \sigma_{av}(\Delta\omega)=\frac{1}{\Delta\omega}
\int_{0}^{\Delta\omega}\sigma(\omega) d\omega\]
\noindent
where $\Delta\omega$ is chosen four to five times larger than the mean finite size
gap of the system as estimated by the ratio of the bare bandwidth and the total number
of eigen values.

\begin{figure}[!t]
\centering
\includegraphics[scale=0.33]{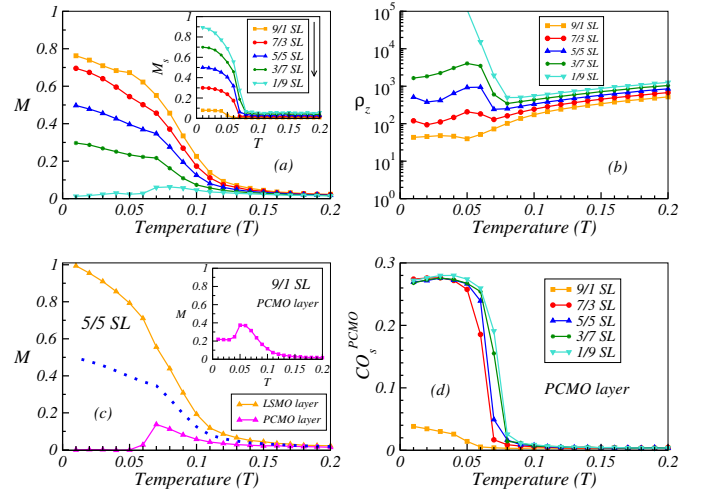}
\caption{
\label{fig03}
Magnetic and transport properties of superlattices at half-doping $n = 0.5$:
(a) Temperature evolution of the magnetization for different superlattices
shows that the total magnetization decreases with increase of the PCMO layer
width.  Inset: Staggered magnetization averaged over individual planes of the
SLs, increases with PCMO layer width $w_P$. (b) $\rho_{z}$ vs $T$ plots of different
superlattices show that the metal-insulator transition temperature ($T_{MI}$)
decreases with the increase of LSMO layer width. (c) The $M$ vs $T$ of individual
LSMO and PCMO layers of 5/5 SL depicts that the LSMO (PCMO) layers have saturation
(vanishingly-small) magnetization at low temperatures. The magnetization of the
whole 5/5 SL is re-plotted for comparison using dotted line. The magnetization 
in the PCMO layers of 9/1 SL is also shown in the inset. (d) Variation of
staggered charge order average over the planes in PCMO layers $CO_{s}^{PCMO}$
vs $T$ for different superlattices shows that the onset of charge ordering
temperature $T_{CO}^{PCMO}$ increases with the increase of PCMO layers width.
The magnitude of $CO_{s}^{PCMO}$ gets saturated at low temperatures, except
for 9/1 SL ($w_{P} = 1$). Legends are same in panels (a), (b) and (d).
}
\end{figure}


\section{Magnetotransport properties of the LSMO/PCMO superlattices at half-doping}

We first estimate the magnetization profile of different superlattices for $n = 0.5$
case as shown in Fig.~\ref{fig03}(a). The onset temperature of the ferromagnetic
ordering of the SLs decreases with the increase of the PCMO layer width $w_P$ i.e.
as we go from 9/1 SL to 1/9 SL as indicated by an arrow near the legend box in
Fig.~\ref{fig03}(a). The magnetization value of the superlattice system decreases
with the increase of the PCMO layer width $w_P$ at low temperatures. In fact, the
magnetization $M$ cease to exist for 1/9 SL at low temperatures. The staggered
magnetization $M_{s}$ averaged over individual planes of the system is plotted
in the inset of Fig.~\ref{fig03}(a), shows that the ground state is dominated by
the antiferromagnetic correlations for 1/9 SL. Now, we calculate the out-of-plane
resistivity ($\rho_z$) for all five superlattices as shown in Fig.~\ref{fig03}(b)
to analyze the correspondence between magnetic and transport properties. Interestingly,
9/1 SL remains more or less metallic at very low temperatures due to the dominant
ferromagnetic nature of the SL system. A metal-insulator transition (MIT) sets up
in an intermediate temperature ($T_{MIT} = 0.07$) for 7/3 and 5/5 SLs. The $T_{MIT}$
increases with increase of the PCMO layer width $w_P$. So, the PCMO layer dominates
the transport properties of the SL at larger $w_P$ and as a result 1/9 SL system is
strongly insulating at low temperatures. The combined result of magnetization and
resistivity of SL systems show that the 1/9 SL behaves as an AF-I system whereas
the 9/1 SL behaves as a FM-M system similar to the bulk LSMO. Overall, the SL
system transforms from a FM-M to AF-I phase with increase of the PCMO layer width
$w_P$. Our calculated magnetotransport results of LSMO/PCMO superlattices at $n = 0.5$
are in good agreement with experimental results~\cite{Nakamura,Wadati}.

\begin{figure}[!t]
\centering
\includegraphics[scale=0.33]{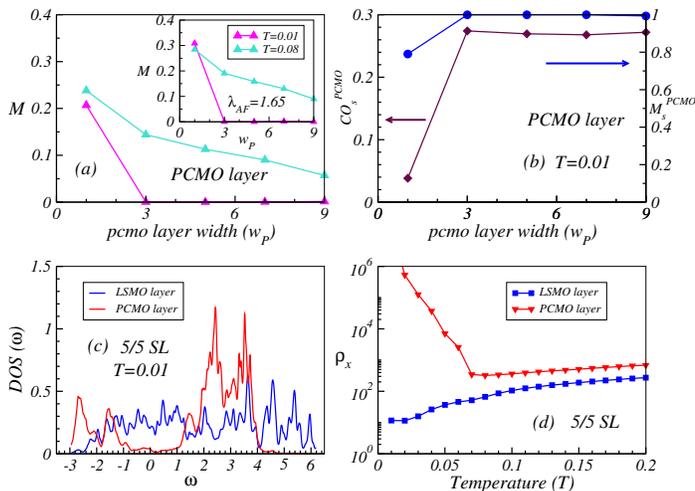}
\caption{
\label{fig04}
(a) Induced Magnetization $M$ in PCMO layer vs PCMO layer width $w_{P}$ plot at
high temperature $T = 0.08$ shows that the induced magnetization decreases
monotonically with the PCMO layer width. At low temperature $T = 0.01$ the
induced magnetization in the PCMO layers vanishes except for $w_{P} = 1$. 
Induced magnetization in PCMO layer vs $w_{P}$ is also plotted for smaller
$\lambda_{AF}$ ($\lambda_{AF} = 1.65$) in the inset. Legends are same in inset
and main figure. (b) Staggered magnetization of PCMO layers $M_{s}^{PCMO}$ nicely
follows the staggered charge ordering $CO_{s}^{PCMO}$ at low temperatures as we
increase the width of the PCMO layer. (c) Density of states of individual LSMO
and PCMO layers for 5/5 SL at low temperature $T = 0.01$ confirms that LSMO
layers have finite density of states at the Fermi level whereas PCMO layers
depicts a gap at the Fermi level. Fermi level is set at $\omega = 0$.
(d) Transverse resistivity $\rho_{x}$ vs $T$ of individual LSMO and PCMO 
layers display that the LSMO (PCMO) layers remains in metallic (insulating)
states at low temperatures. All the calculations are performed at electron density
$n = 0.5$.
}
\end{figure}

  
To further understand the analogy between the observed magnetizations and
the metal-insulator transitions in Figs.~\ref{fig03}(a) and (b), particularly
at intermediate $w_P$ values, we present the average magnetization of the
individual LSMO and PCMO layers in Fig.~\ref{fig03}(c) for 5/5 SL. The total
magnetization is replotted in same figure. It is clear that the PCMO layer
magnetization starts to decrease below the temperature $T\sim0.07$ and vanishes
at very low temperatures. This decrease in PCMO layer magnetization
is associated with the formation of charge ordering in the PCMO layer at that
temperature. We plot the staggered charge ordered $CO_{s}^{PCMO}$ vs temperature
of PCMO layers for 5/5 SL in Fig.~\ref{fig03}(d). The onset of charge ordering
at $T \sim 0.07$ negates the induced ferromagnetic moments in PCMO layer below
$T = 0.07$. In fact, the slope of the magnetization curve [see Fig.~\ref{fig03} (a)]
of total 5/5 SL changes at $T = 0.07$ which matches well with the charge ordering
temperature. This temperature also coincides with the metal-insulator transition
presented in Fig.~\ref{fig03}(b). All these coherent results confirm that the main
contribution to the magnetization of 5/5 SL comes only from the LSMO layer at
low temperatures. The charge ordering of PCMO layers at low temperatures is also
apparent for 7/3, 3/7 and 1/9 SLs [see Fig.~\ref{fig03}(d)] that prohibits the
magnetization in PCMO layer. The diminishing magnetization of PCMO layers
decreases the magnetization of the overall superlattice system with the increase
of PCMO layer width as shown in  Fig.~\ref{fig03}(a). But the charge ordering
temperature and strength of charge ordering at low temperatures in PCMO layer
remains comparatively small for $w_P$ = 1. As a result the induced
magnetization in antiferromagnetic PCMO layer survives at low temperatures for
9/1 SL as shown in inset of Fig.~\ref{fig03}(c).

Next, we plot the induced magnetization in the antiferromagnetic PCMO layer of
the superlattice systems for different width of the PCMO layer at two different
temperatures [i.e. at low ($T = 0.01$) and high ($T = 0.08$) temperatures] in
Fig.~\ref{fig04}(a) to compare our results elaborately. Temperature $T = 0.08$
is below the ferromagnetic $T_C$ of the LSMO layer but is larger than the
charge ordering temperature of PCMO layers presented in Fig.~\ref{fig03}(d).
The induced magnetization in PCMO layers is finite for both $T = 0.01$ and
$T = 0.08$ for $w_P$ = 1. At high temperatures ($T = 0.08$), above the onset
of the charge ordering in PCMO layer, the induced magnetization persists up
to larger $w_P$. At very high value of $w_P$ the magnetization in PCMO layer
will approach the bulk limit. Does the induced magnetization profile changes
drastically if we increase the bandwidth of the material? To answer this we
show the induced magnetization in PCMO layer using $\lambda_{AF}$ = 1.65
(smaller $\lambda_{AF}$ corresponds to larger bandwidth or vice versa~\cite{Pradhan3})
in the inset of Fig.~\ref{fig04}(a). Our qualitative results remain almost
unchanged.

Now we turn back to $\lambda = 1.7$ case. The induced magnetization in PCMO
layer suddenly decreases with the increase of PCMO layer width at $T = 0.01$,
unlike at $T = 0.08$ where magnetization decreases rather slowly. There is a
one-to-one correspondence between the induced magnetization and the charge
ordering strength at $T = 0.01$ in PCMO layers as shown in Fig.~\ref{fig04}(b).
As we mentioned above the strong charge ordered states for all SLs except 9/1
SL prohibits the induced magnetization in PCMO layers. The staggered
magnetization in PCMO layers [see Fig.~\ref{fig04}(b)] also follows the same
trend to that of $CO_{s}^{PCMO}$ at low temperatures. These results emphasize
that the PCMO layers (except for $w_P$ = 1) in LSMO/PCMO SLs are in charge
ordered antiferromagnetic state at low temperatures. To reveal the insulating
nature of the PCMO layers we plot the layer resolved density of states (DOS) of
the 5/5 SL [see Fig.~\ref{fig04}(c)] at $T = 0.01$ for both LSMO and PCMO layers.
The Fermi level is set at $\omega$ = 0. The DOS is finite at Fermi level for
the LSMO layers whereas DOS is gapped at Fermi level for the PCMO layer. This
indicate that LSMO (PCMO) layers are in metallic (insulating) state. The
transverse resistivity $\rho_x$ of each layer with temperature (shown in
Fig.~\ref{fig04}(d)) further corroborates this scenario.

\begin{figure}[!t]
\centering
\includegraphics[scale=0.33]{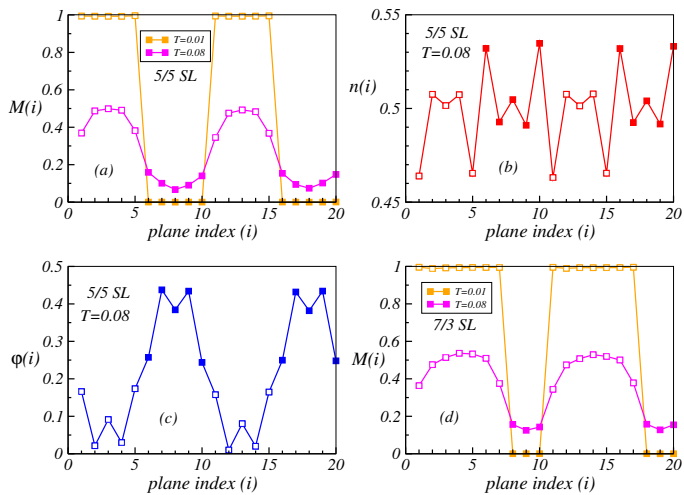}
\caption{
\label{fig05}
Individual plane average magnetization ($M(i)$) vs plane index ($i$) at low
temperature ($T = 0.01$) and high temperature ($T = 0.08$) for (a) 5/5 SL
($w_{P}=5$), (d) 7/3 SL ($w_{P}=3$). Plane averaged (b) electron density ($n(i)$)
vs plane index ($i$), (c) LRC potential ($\varphi(i)$) vs plane index ($i$) for
5/5 SL ($w_{P} = 5$) at $T = 0.08$. Opened (filled) symbol corresponds to LSMO (PCMO) 
planes.
}
\end{figure}

Next, we plot the plane-wise magnetization [M(i)] profile for 5/5 SL at
$T = 0.01$ in Fig.~\ref{fig05}(a). The magnetization is saturated for each LSMO
plane whereas the induced magnetization is negligible in all PCMO planes. This
is obvious as magnetization in PCMO layer is negligibly small as we discussed in
Fig.~\ref{fig03}(c). It is apparently clear from our analysis that the induced
magnetization in PCMO layer persists in a small temperature window which is
above the charge ordering temperature $T_{CO}$ in PCMO layer and below the FM
$T_C$ of LSMO layer except for $w_P$ = 1. Does each plane of PCMO layers gets
magnetized equally in this temperature window? To answer this we have plotted
plane-wise magnetization [M(i)] vs plane index ($i$) of 5/5 SL at $T = 0.08$ in
Fig.~\ref{fig05}(a). All the PCMO planes have finite magnetization which justify
the magnetization of PCMO layer vs temperature curve shown in Fig.~\ref{fig03}(c).
Magnetization in interfacial planes is comparatively larger than the inner planes
in PCMO layers at this temperature regime. On the other hand, the magnetization
decreases as we move from inner planes to interfacial planes in LSMO layer.
This is due to the average electron density variation as a result of charge
transfer across the interfacial planes from LSMO to PCMO, which is evident from the
plane-wise electron density profile [$n(i)$ vs $i$] plotted in Fig.~\ref{fig05}(b).
It also shows that the charge transfer occurs mainly at the interfacial planes
whereas the electron density of inner planes of both LSMO and PCMO layers remains
very close to the initial electron density, i.e. $n$ = 0.5. Furthermore, the
associated LRC potential profile, shown in Fig.~\ref{fig05}(c), is almost
symmetric and a potential gradient across the interface is observed. This
confirms that the amount of charge transfer occurs across the interface is
very much controlled due to LRC potentials. Hence, the decrease of electron
density from the interfacial LSMO layer, due to the charge transfer, leads to
the decrease of the magnetization value at the interfacial LSMO layers,
although minimal, as compared to the inner planes of LSMO layer
[see Fig.~\ref{fig05}(a)].

In addition, plane-wise magnetization profile of the 7/3 SL at two different
temperatures ($T$ = 0.01 and 0.08) are also shown in Fig.~\ref{fig05}(d). They
exhibit very similar characteristics as in 5/5 SL. These magnetization profiles
at low temperature $T = 0.01$ [see Figs.~\ref{fig05}(a) and (d)] reveal that
the magnetization of LSMO layer is saturated but the PCMO layer has negligible
magnetization and as a result the SL system displays {\it inhomogeneous}
phase-separated-like state. On the other hand the LSMO and PCMO layers of the SLs
have finite magnetization at high temperatures ($T = 0.08$) and the systems
behave relatively {\it homogeneous}. The {\it homogeneous} ({\it inhomogeneous})
phase of the superlattices at high (low) temperatures matches well with the
reported experimental results performed on this type of manganite
superlattices~\cite{Nakamura,Wadati}. Overall, it seems quite strenuous to
destabilize the CO state (or in other terms, induce magnetization) in the PCMO
layer at $n$ = 0.5 by varying the interfacial mechanisms through variation of
PCMO layer thickness at low temperatures, mainly due to the robustness of the
CO. Thus, the next step is analyze LSMO/PCMO superlattices at off-half-doping.

\begin{figure}[!t]
\centering
\includegraphics[scale=0.33]{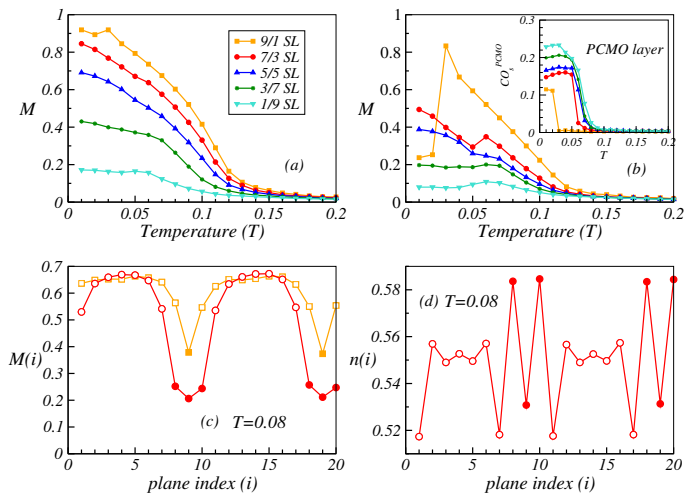}
\caption{
\label{fig06}
Magnetic properties at off-half-doping $n$ = 0.55: (a) Magnetization $M$ vs
$T$ plots for various superlattices depict that the transition temperature
($T_{C}$) as well as saturation magnetization at low temperature decreases
with increase of the PCMO layers width $w_{P}$. (b) Temperature dependence
of the magnetization of PCMO layers are also plotted for different SLs. 
Inset: Average $CO_{s}^{PCMO}$ vs $T$ reveals that the change ordering
temperature in the PCMO layers ($T_{CO}^{PCMO}$) and saturation value of
$CO_{s}^{PCMO}$ at low temperatures increases with increase of PCMO layers
width ($w_{P}$). Legends used in (a) and (b) are same. (c) Plane averaged 
magnetization ($M(i)$) vs plane index $i$ for 9/1 SL ($w_{P} = 1$) and 7/3 SL
($w_{P}=3$) show that the induced magnetization in PCMO layer for 9/1 SL is
comparatively larger than 7/3 SL at $T = 0.08$. (d) Plane averaged electron
density ($n(i)$) vs plane index $i$ for 7/3 SL depicts the charge transfer
occurs at interface. 
}
\end{figure}

\begin{figure}[!t]
\centering
\includegraphics[scale=0.33]{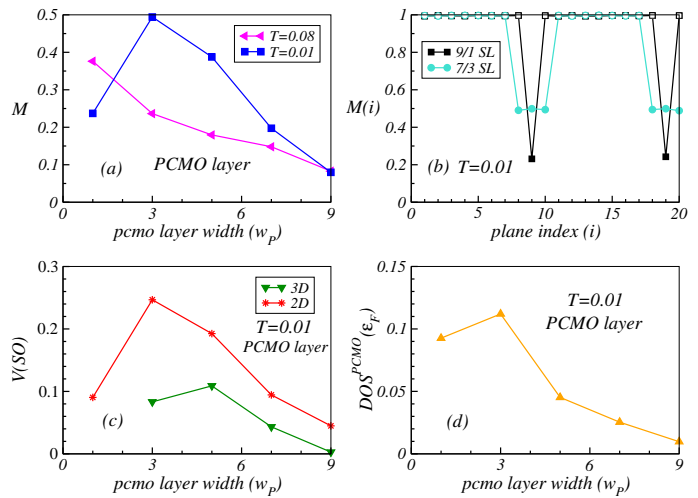}
\caption{
  \label{fig07}
Magnetotransport properties of LSMO/PCMO superlattices at $n = 0.55$: 
(a) Induced magnetization in PCMO layers with the width of PCMO layers
($w_{P}$) displays the nonmonotonic behavior at $T = 0.01$ in contrast to the
the high temperatures ($T = 0.08$). At $T = 0.01$ the magnetization of the PCMO
layers achieves optimum value for $w_{P} = 3$. (b) Plane averaged magnetization
($M(i)$) vs plane index ($i$) for 9/1 and 7/3 SLs at $T = 0.01$ show that the
induced ferromagnetic moments in PCMO layer are enhanced for $w_P = 3$.
(c) 2D and 3D volume fraction of the ferromagnetic nanocluster (V(SO)) in the 
PCMO layers vs PCMO layer width $w_{P}$ are plotted at $T = 0.01$. 2D volume
fraction perfectly follows the magnetization profile of the PCMO layers as in
(a). For more details please see the text. (d) Density of states of
the PCMO layers at the Fermi level ($DOS^{PCMO}(\varepsilon_{F})$) also
shows nonmonotonic behavior similar to magnetization of the PCMO layers
at $T = 0.01$.
}
\end{figure}

\section{Magnetotransport properties of the superlattices at off-half-doping}

As we mentioned earlier, the AF phase at off-half-doping is not a perfect charge
and orbital ordered (CO-OO) phase; the CO-OO is inherently somewhat diluted. Now
the question arises: is it possible to destabilize the CO-OO state further in the
off-half doped PCMO by tuning its layer thickness ($w_P$) in LSMO/PCMO superlattices?
To answer this question, we study the similar kind of SLs with off-half doped
PCMO, especially at $n$ = 0.55. In order to interpret the systematics of induced
magnetic moments in PCMO layers and its effect on transport properties we analyze
LSMO/PCMO superlattices at $n = 0.55$ for various width of PCMO layers. We plot
the magnetization vs temperature in Fig.~\ref{fig06}(a) at $n = 0.55$. The onset
temperature of the ferromagnetic ordering of SLs decreases with the increase of
PCMO layer width $w_P$, similar to $n = 0.5$ case. The low temperature magnetization
value of the SLs also decreases with $w_P$. The small peak at $T = 0.03$ observed
for 9/1 SL is due to the sudden decrease in the magnetization of the PCMO layer
as shown in Fig.~\ref{fig06}(b). This fall in PCMO layer magnetization at
$T = 0.03$ is due the commencement of charge ordering [see the inset of
Fig.~\ref{fig06}(b)] in the PCMO layer. The magnetization of the PCMO layer in
7/3 SL also decreases at $T = 0.06$ because of the onset of charge ordering in
PCMO layer, but the magnetization increases again below $T = 0.05$. In fact, the
charge ordering is observed for all SLs (charge ordering temperature increases
with width of PCMO layer) but the abrupt change in the magnetization of the
embedded PCMO layer near the charge ordering temperature is more prominent for
thinner PCMO layers only. The induced magnetization in the thinner PCMO layers
are relatively large at the time of the onset of charge ordering and as a result the
drop is more prominent. Also, the induced magnetization sustains the impact of
charge ordering for larger values of $w_P$.

Next, to reveal the induced moment across the planes, we plot the plane-wise
magnetization of 9/1 and 7/3 SLs at $T = 0.08$ (just above the charge ordering
temperature of PCMO layers in 7/3 SL) in Fig.~\ref{fig06}(c). For visual impact
we have shifted the result of 9/1 SL by one plane which is allowed due to
the imposed periodic boundary conditions in our SL system.
The induced magnetization in each PCMO plane is apparent from this plot. It is
also clear that the magnetization of the interfacial LSMO planes are relatively
smaller than the inner LSMO planes similar to $n = 0.5$ case. The plane-wise
electron density profiles plotted for 7/3 SL in Fig.~\ref{fig06}(d) shows that
the charge transfer occurs across the interfaces, from LSMO to PCMO and as a
result the interfacial magnetizations in LSMO layer decreases.

We plot the average induced magnetization in PCMO layer at $T = 0.08$ in
Fig.~\ref{fig07}(a). This induced magnetization in PCMO layer decreases
monotonically with the increase of PCMO layer width, similar to the results
we obtained at high temperatures for $n = 0.5$ case. Please note that
$T = 0.08$ is higher than the charge ordering temperatures of the embedded
PCMO layers in our SLs. At low temperatures charge ordering is observed in
the PCMO layers for all our SLs. But the strength of the charge ordering in
PCMO layer increases with $w_P$ [as shown in inset of Fig.~\ref{fig06}(b)].
It is important to note that the strength of the charge ordering in PCMO
layer for $n = 0.55$ case is comparatively weaker to that of $n = 0.5$ case.
The weakened charge ordering at intermediate $w_P$ values give rise to the
coexistence of induced magnetization and charge ordering in the PCMO layer
at low temperatures [Figs.~\ref{fig06}(a) and (b)] which was not observed
in $n = 0.5$ case.

\begin{figure}[!t]
\centering
\includegraphics[scale=0.33]{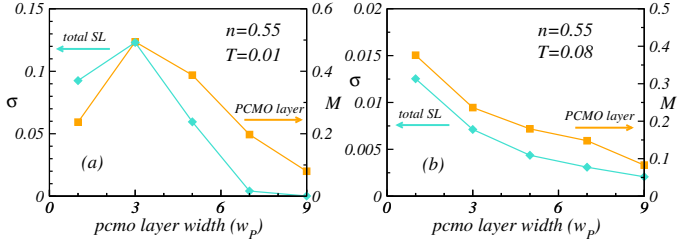}
\caption{
\label{fig08}
One-to-one correspondence between magnetic and transport properties ($n = 0.55$): 
(a) Conductivity $\sigma$ vs $w_{P}$ exhibits nonmonotonic behavior similar
to the induced magnetization in PCMO layers at $T=0.01$. (b) At $T = 0.08$,
$\sigma$ vs $T$ depicts that the conductivity diminishes monotonically with
increase of $w_{P}$ which is also similar to the magnetization profile of
PCMO layer.}
\end{figure}

The induced magnetization in the PCMO layers at low temperature $T = 0.01$ for
SL for $n = 0.55$ SL, in contrary to $n = 0.5$ SLs, remains finite for all $w_P$
values studied in this work [see Fig.~\ref{fig07}(a)]. Interestingly the induced
magnetization in PCMO layers shows nonmonotonic behavior with $w_P$. The optimum
magnetization occurs for $w_P = 3$ i.e., for 7/3 SL. The plane-wise magnetization
profiles are shown for $w_P = 1$ (9/1 SL) and $w_P = 3$ (7/3 SL) in
Fig.~\ref{fig07}(b) at $T = 0.01$. Here also, we have shifted the result
of 9/1 SL by one plane [as in Fig.~\ref{fig06}(c)]. The plane-wise magnetization
profile shows that the induced magnetization in individual PCMO planes
are smaller in 9/1 SL as compared to that of 7/3 SL. To further probe the
induced magnetism in PCMO layer we calculate the volume fraction of the
ferromagnetic clusters [V(SO)] inside the PCMO layers. The low temperature 3D
volume fraction shows that the fraction of ferromagnetic clusters decreases
considerably for large $w_P$ [Fig.~\ref{fig07}(c)] and as a result the
magnetization of PCMO layer will approach the bulk PCMO limit at very large
$w_P$. But, at intermediate values of $w_P$, we observed that a considerable
amount of ferromagnetic clusters are present in PCMO layers. Unfortunately
V(SO) of PCMO layer can not be calculated for $w_P = 1$. To include the
ferromagnetic volume fraction of $w_P = 1$ we evaluate the 2D volume fraction
(calculated for each plane and averaged over the number of planes) of the PCMO
layers. The low temperature 2D volume fraction shows nonmonotonic behavior with
$w_P$ similar to the induced magnetization profile shown in Fig.~\ref{fig07}(a).
The induced magnetization in PCMO layers will most probably affect on the
transport properties of the SLs. Hence in our next step we calculate the
density of states at the Fermi level of the PCMO layers $DOS^{PCMO}(\epsilon_{F})$
to ascertain the behavior of transport properties more closely at low
temperatures ($T = 0.01$) for different SLs as shown in Fig.~\ref{fig07}(d).
$DOS^{PCMO}(\epsilon_{F})$ for different $w_P$ values also exhibit a nonmonotonic
behaviors similar to the induced magnetization profile. All these coherent
results indicate that the conductivity will also follow a nonmonotonic path
with variation of PCMO layer width.

In fact, the out-of-plane conductivity $\sigma$ (see Fig.~\ref{fig08}(a)) of
the superlattices follows a similar nonmonotonic characteristics at $T = 0.01$
with optimum conductivity at $w_P = 3$. This shows that induced magnetization in
the PCMO layers indeed smoothen the conducting paths between the ferromagnetic
LSMO channels by reducing the scattering processes during the electrical conduction.
As a result, the conductivity follows the induced magnetization profile of the
embedded PCMO layers in the SL structures. On the other hand, the conductivity
$\sigma$ decreases monotonically with $w_P$ at high temperatures which is
analogous to the induced magnetization profile in the embedded PCMO layers
[see Fig.~\ref{fig08}(b)]. Overall, our calculations establish a one-to-one
correspondence between the magnetic and transport properties in the LSMO/PCMO
superlattices. The nonmonotonic characteristics of the magnetotransport
properties with variation of $w_P$ matches with the reported experimental
results~\cite{Nieb1,Nieb2}.

\begin{figure}[!t]
\centering
\includegraphics[scale=0.33]{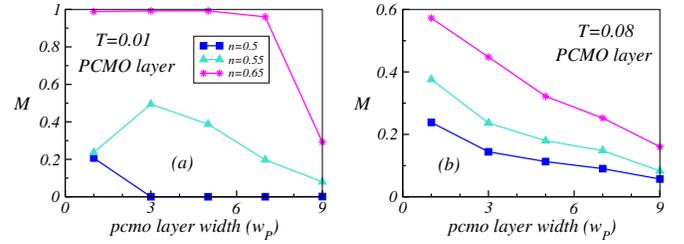}
\caption{
\label{fig09}
Comparison of magnetization in PCMO layers for three different densities $n = 0.5,
0.55$, and $n=0.65$ at (a) $T=0.01$, (b) $T=0.08$. 
}
\end{figure}

Finally, we want to show that the induction of magnetization is more easier
as we go even further away from half-doping. For this, we calculate the
induced magnetization profile in the PCMO layers for different SLs using
$n = 0.65$ at low ($T = 0.01$) and high ($T = 0.08$) temperatures. The
induced magnetization profile in the PCMO layers of the superlattices for
three different densities $n = 0.5, 0.55$ and $0.65$ are shown in
Fig.~\ref{fig09}(a) and ~\ref{fig09}(b) at low ($T = 0.01$) and high
($T = 0.08$) temperatures, respectively for comparison. For $n=0.65$ case
the induced magnetization in the PCMO layers remains saturated up to
$w_P = 7$ and then start to recede (see Fig.~\ref{fig09}(a)). This indicates
that the entire superlattice convert to a ferromagnet for $w_P \le 7$. So,
it is easier to magnetize weakened charge ordered antiferromagnetic layer
at $n = 0.65$. In the contrary, although induced magnetization for $n=0.65$
case is comparatively large to that of $n = 0.5$ and $n = 0.55$ cases, it
decreases monotonically starting from $w_P = 1$ at high temperatures as
shown in Fig.~\ref{fig09}(b).

Overall, our result at $n = 0.65$ does not show any nonmonotonicity with
variation of $w_P$. If one wants to compare our results one-to-one with
experiments at $n \sim 0.65$~\cite{Nieb1} then we agree that nonmonotonicity
of the magnetization in the antiferromagnetic layer at low temperatures is not
reproduced exactly in our calculations. This may be due to the disorder effect
in thinner PCMO layers that we have neglected in our calculations. Thin PCMO
layers and interfacial planes (of both LSMO and PCMO layers) are likely to
be more vulnerable to disorder due to strain effects and as a result the
magnetization in thin PCMO likely to decrease in presence of disorder~\cite{Pradhan}.
At the same time it is noteworthy to recover the nonmonotonic nature of
the induced magnetization in PCMO layer at $n = 0.55$ without taking any
disorder effect into account.

\section{Conclusions}
In this paper, we have studied the magnetotransport properties of manganite based
LSMO/PCMO like FM/AF superlattice systems using two-orbital double exchange model
by incorporating superexchange interaction and Jahn-Teller lattice distortions. We
focus on the induced ferromagnetism in PCMO layer at half-doping ($n$ = 0.5) and
off-half-doping ($n$ = 0.55) cases that give rise to intriguing phase coexistence
in FM/AF SL systems. The induced ferromagnetic moments in the antiferromagnetic
PCMO layer decreases monotonically with increasing the antiferromagnetic layer
width at high temperatures (just above the charge ordering temperature of PCMO
layers but, below the ferromagnetic $T_C$ of LSMO) for both half-doping and
off-half-doping scenarios. As we decrease the temperature the induced ferromagnetic
moments in PCMO layer disappears or decreases considerably for half-doping. This
is due to the onset of charge ordering in antiferromagnetic PCMO layers. Our
calculations shows that the metal-insulator transition temperature of the
LSMO/PCMO superlattices increases with increase of the PCMO layer width $w_P$,
similar to the experiments. We outline the underlying one-to-one correspondence
between magnetic and transport properties. However, contrasting features appear
at off-half-doping regime ($n$ = 0.55). Our calculations comprehensively show
that it is easier to induce ferromagnetic moment in weakened charge ordered state
at $n$ = 0.55 even at low temperatures. Interestingly, the induced magnetization
in PCMO layer and the conductivity of the total superlattice system shows
nonmonotonic behavior with increase of $w_P$ in agreement with experiments.

\begin{center}
\textbf{ACKNOWLEDGEMENTS}
\end{center}

We acknowledge use of the Meghnad2019 computer cluster at SINP.


\end{document}